\documentclass[12pt]{article}
\usepackage{geometry} 
\pdfoutput=1
\usepackage[]{graphicx}  
\usepackage{caption}
\usepackage{natbib}
\usepackage{mathtools}
\usepackage{authblk}
\title{The pattern and distribution of deleterious mutations in maize}
\author[1]{Sofiane Mezmouk\thanks{smezmouk@ucdavis.edu}}
\author[1,2]{Jeffrey Ross-Ibarra\thanks{rossibarra@ucdavis.edu}}
\affil[1]{Department of Plant Sciences, University of California Davis}
\affil[2]{Center for Population Biology and Genome Center, University of California Davis}

\date{}

\begin{document}
\maketitle

\begin{abstract} 
Most non-synonymous mutations are thought to be deleterious because of their effect on protein sequence.  
These polymorphisms are expected to be removed or kept at low frequency by the action of natural selection, and rare deleterious variants have been implicated as a possible explanation for the ``missing heritability'' seen in many studies of complex traits. Nonetheless, the effect of positive selection on linked sites or drift in small or inbred populations may also impact the evolution of deleterious alleles. 
Here, we made use of genome-wide genotyping data to characterize deleterious variants in a large panel of maize inbred lines.  
We show that, in spite of small effective population sizes and inbreeding, most  putatively deleterious SNPs are indeed at low frequencies within individual genetic groups. 
We find that genes showing associations with a number of complex traits are enriched for deleterious variants. 
Together these data are consistent with the dominance model of heterosis, in which complementation of numerous low frequency, weak deleterious variants contribute to hybrid vigor.
\end{abstract}

\newpage
\section*{Introduction}

The effect of new mutations on organismal fitness is not well understood, but both theoretical considerations \citep{Fisher1930} and empirical estimates \citep{Joseph2004} suggest that most new mutations are deleterious and only a small minority are beneficial. Strongly deleterious mutations are expected to be kept at low frequencies by natural selection, whereas weakly deleterious alleles may be effectively neutral \citep{Ohta1973, Kimura1983} and subject to the effects of genetic drift \citep{Lynch1990,Lande1994,Whitlock2003}. In addition to selection and drift, a number of other factors such as mating system and recombination rate also impact the evolution of deleterious alleles. Selfing species and inbreeding within populations will expose lethal mutations to selection faster than in an outcrossing population \citep{Wang1999,Glemin2003}, but weakly deleterious mutations  can nonetheless be maintained at moderate frequencies, even in the presence of gene flow between populations \citep{Whitlock2000}. Moreover, in genomic regions with low levels of recombination, selection against deleterious mutations will be less effective \citep{Charlesworth1993fp} and the potential exists for deleterious mutations to rise to high frequency due to the effects of linked selection on beneficial mutations \citep{Felsenstein1974, Hill1966, Chun2011}. 
 
Deleterious alleles may play an important functional role in affecting the phenotype of traits of interest, and complementation between haplotypes carrying different deleterious alleles may explain much of the observation of hybrid vigor or heterosis \citep{Charlesworth2009}. 
In human disease studies, significant correlation was observed between the deleterious predictions of SNPs and their association with cancer \citep{Zhu2004}; predicted rare deleterious SNPs were also shown to be involved in common diseases \citep{Cohen2004,Smigrodzki2004}.
Furthermore, rare deleterious SNPs have gained interest due to their potential role in explaining quantitative trait variation \citep{Gibson2012}, especially in populations that have experienced recent growth \citep{LohmuellerArXiv} .

Evaluating the abundance and frequency of deleterious mutations is thus of considerable interest and has been investigated in a wide range of species. These analyses have varied in terms of the percentage of non-synonymous sites estimated to be deleterious, from 3\% in bacterial populations \citep{Hughes2005} to 80\% in the human genome \citep{Fay2001}. They have also shown that recently bottlenecked populations may have a higher abundance of deleterious sites and that heterozygosity at deleterious SNPs is lower than at synonymous SNPs \citep{Lohmueller2008}.
Among plants, deleterious variants have been investigated in detail in only a few cases. \citet{Gossmann2010} found that most new mutations in plants are strongly deleterious, with only ~25\% acting as effectively neutral. \citet{Cao2011} show that the abundance of deleterious variants correlates with effective population size of \emph{Arabidopsis thaliana}, but comparisons of purifying selection in several species of tomato also implicate a role for environmental differences as well \citep{Tellier2011}. In natural populations of \emph{Arabidopsis thaliana}, selection appears to act to maintain variants that are locally adaptive but deleterious elsewhere \citep{Fournier-Level2011}, whereas positive selection on domestication genes may have increased the abundance of deleterious variants in domesticated genomes such as rice \citep{Gunther2010, Lu2006}. While these studies have provided insight into the evolutionary fate of deleterious mutations, we still understand relatively little about the role of deleterious variants in affecting pheotypic traits. 

Maize (\emph{Zea mays}) is a worldwide economically important cereal with the highest yield and one of largest crop cultivated area (FAO statistics); it is also a great model for basic and applied research \citep{Strable2009}. Maize was traditionally cultivated 
as open pollinated populations (landraces) but, after the first documented observations of hybrid vigor in this species \citep{East1908,Shull1908}, inbred lines were developped and structured into heterotic groups that maximize inter-group combining ability. 
The transition from heterozygous populations to inbred, strongly structured heterotic groups, makes maize inbred lines of interest for analyzing the distribution and frequency of deleterious mutations. Furthermore, high observed values of hybrid vigor or heterosis in maize hybrids makes it a great model for testing the effects of deleterious mutations and their contribution to heterosis.  The dominance model of heterosis posits that inbred lines are homozygous for a number of recessive deleterious alleles and that crosses between inbreds carrying different complements of deleterious alleles will result in heterozygous progeny with higher fitness than either parent.

The aim of the current study was to make use of the availability of the maize genome sequence, high density single nucleotide polymorphisms (SNPs) and phenotypic data for a large sample of inbred lines and hybrids to (1) carry out a genome wide scan for deleterious mutations, (2) analyze their distribution across the genome and within different genetic groups and (3) test for enrichment of these loci in the results of genome wide association mapping. Our results showed that maize inbred lines are segregating for a large number of predicted deleterious variants (20 to 40 \,\% of protein coding SNPs were predicted to have a deleterious allele), and that these alleles are generally at very low frequencies with few fixed differences observed among different genetic groups. Genome-wide association analysis of hybrid vigor finds little evidence for enrichment of individual deleterious SNPs, but significant enrichment for genes containing deleterious SNPs, suggesting a meaningful role for dominance and complementation in explaining observations of hybrid vigor. 

\section*{Materials and methods}
\subsection*{Plant material and phenotypic data}

We utilized phenotypic data published in \citet{Flint-Garcia2005} for 247 maize inbred lines (see supplemental data for a list of inbred lines). Each inbred line was crossed to the stiff-stalk inbred B73 (population A) and both the inbred lines and their B73-hybrids were evaluated in three environments in 2003 \citep{Flint-Garcia2009}. A subset of 102 inbreds were additionally crossed to both B73 (population B1) and Mo17 (population B2) and evaluated in a single environment in 2006 \citep{Flint-Garcia2009}. Supplemental Table~\ref{Traits} lists the analyzed traits which are detailed in \citet{Flint-Garcia2009}.

The panel structure was previously analyzed and inbred lines were attributed to different subpopulations \citep{Flint-Garcia2005}. For the main temperate inbred lines, these subpopulations corresponds to the different heterotic groups.

\subsection*{Genotypic data}

We made use of genotypic data from \citet{Larsson2013} for the full set of 247 lines. The latter were genotyped using the genotyping-by-sequencing approach \citep[GBS;][]{Elshire2011}, resulting in a total of 437,650 SNPs that were partially imputed. Of these SNPs, 127,994 mapped to protein coding sequences representing 123,289 codons in 21,064 genes. The median (mean) percentage of missing data per SNP, including triallelic sites, was 1.06\% (2.52\%), while the percentage of heterozygous sites was 1.08\% (2.52\%). Only 4.5\% of SNPs had more than 10\% missing data (Supplemental Figure~\ref{figureS1}-A), and 0.18\% had more than 10\% heterozygous genotypes (Supplemental Figure~\ref{figureS1}-B).

We estimated error rates by first comparing our genotyped inbred B73 to the B73 reference genome, then by comparing all our genotypes to those from 7,225 overlapping SNPs on the maize SNP50 bead chip \citep{Cook2012}.  
Compared to the reference genome, our B73 genotype after imputation differed (alternative homozygote allele) at 1.75\% of SNPs, and across all lines our genotypes differed at a median (mean) rate of 1.83\% (4.62\%) from the maize SNP50 data \citep{Cook2012}.   

\subsection*{Statistical analyses}

\subsubsection*{SNP annotation and analyses}
 The first transcript of each gene in the B73 5.b filtered gene set was used to annotate SNPs as synonymous and non-synonymous with the software polydNdS from the analysis package of libsequence  \citep{Thornton2003}. The deleterious effects of amino acid changes were then predicted for proteins derived from the first transcript of each gene with both the SIFT \citep{Ng2003, Ng2006} and MAPP \citep{Stone2005} software packages.

SIFT uses homologous sequences identified by PSI-BLAST against protein databases to identify conserved amino acids.  
The software provides a scaled score of the putative deleterious effect of a particular amino acid at a position along a protein. 

MAPP predicts deleterious amino acid polymorphisms from a user-defined alignment of protein homologs. It uses the phylogenetic relatedness among sequences and the physicochemical properties of amino acids to quantify the potential deleterious effect of a given amino acid change.  We created alignments for MAPP using three different methods.  First, we made BLASTX comparisons of protein sequences from maize against the TrEMBL database \citep{Boeckmann2003}, retaining all proteins with an e-value $\leq 10^{-40}$ and at least 60\% identity with the query.  Second, we used a reciprocal best BLAST criterion to compare protein sequences of maize against protein sequences from 31 plant genomes (supplemental data)  from Phytozome version 8.0 (\url{http://www.phytozome.net}), retaining the best hit protein from each of the other genomes with a minimum e-value $\leq 10^{-100}$ and  $\geq 70\%$ coverage of the query length. Finally, we made use of a set of syntenic genes from the grasses \emph{Zea mays}, \emph{Sorghum bicolor}, \emph{Oryza sativa} and \emph{Brachypodium distachyon}  \citep{Schnable2012}. For each set of proteins, ClustalW2 \citep{Larkin2007} was used to align the sequences and build a neighbour-joining tree. A custom R script was used to link amino acid positions to SNP positions and to link the amino acid polymorphisms to MAPP and SIFT predictions.

The derived site frequency spectrum was calculated for all protein coding SNPs with \emph{Tripsacum} as outgroup. 
The pattern of haplotype sharing across the genome \citep[PHS statistics;][]{Toomajian2006} was analyzed within each of the tropical, stiff-stalk, non-stiff stalk and mixed subpopulations as defined by \citet{Flint-Garcia2005}. We will refer to these subpopulations as ``genetic groups".

\subsubsection*{Phenotypic data analyses}

Genetic values 
of inbreds and hybrids in population B were taken from \citet{Flint-Garcia2009}. 
Genetic values for population A were estimated from the raw phenotypic data using the model:
\[Y=\mathbf{1}\mu +ZG+\varepsilon \]
where $Y$ is the vector of phenotypic values, $\mu$ is the mean of $Y$, $Z$ is an incidence matrix, $G$ is the vector of fixed individual effects and $\varepsilon$ are the residuals assumed to be $\mathcal{N}(0,\sigma _{\varepsilon }^{2}I)$.

Hybrid vigor for each individual was estimated by both best- and mid-parent heterosis ($BPH$ and $MPH$, respectively):
\[ MPH_{ij}=\hat{G_{ij}}-\frac{1}{2}(\hat{G_{i}}+\hat{G_{j}}) \]
\[ BPH_{min,ij}=\hat{G_{ij}}-min(\hat{G_{i}} ,\hat{G_{j}}) \] 
\[ BPH_{max,ij}=\hat{G_{ij}}-max(\hat{G_{i}} ,\hat{G_{j}}) \]
where $\hat{G_{ij}}$, $\hat{G_{i}}$ and $\hat{G_{j}}$ are the genetic values of the hybrid and its two parents $i$ and $j$. $BPH_{min}$ was used instead of $BPH_{max}$ for days to anthesis, tassel branch count, tassel angle, upper leaf angle and rind penetrometer resistance.

\subsubsection*{Association mapping}

SNP association with the genetic values of the inbred lines were tested using the mixed linear model:
\[\hat{G}=\mathbf{1}\mu + M\vartheta +S\beta +Zu+\varepsilon\]
where $\hat{G}$ is the vector of estimated genetic values for inbred lines, $\mu$ is the mean of $\hat{G}$, $M$ is the tested SNP, $\vartheta$ is the SNP effect, $S$ is the structure covariates estimated  with STRUCTURE software \citep{Pritchard2000} by \citet{Flint-Garcia2005}, $\beta$ is the fixed structure effects, $Z$ is an incidence matrix, $u$ is a random effect vector assumed to be $\mathcal{N}(0,\sigma_{ {\varepsilon}}^{2}K)$ and $\varepsilon$ are the model residuals assumed to be $\mathcal{N}(0,\sigma_{ {\varepsilon}'}^{2}I)$. 
The coancestry matrix $K$ among inbred lines was approximated by an identity by state matrix calculated with the SNPs. Only SNPs with a minor allele frequency $\ge$ 0.05 were used for association mapping tests.

In hybrids, we tested the effect of heterozygosity at a given locus on observed heterosis. 
Each SNP was assigned numerical values corresponding to $0$ if the hybrid is homozygous or $1$ if the hybrid is heterozygous. 
The association mapping tests were thus carried out between heterozygosity at a given locus and hybrid vigor:
\[PH=\mathbf{1}{\mu}'+D\beta +H\vartheta +{\varepsilon }'\]  \\
where $PH$ is either $MPH$, $BPH_{max}$ or $BPH_{min}$, ${\mu}'$ is the mean of $PH$, $D$ is the genetic distance between the tester (B73 or Mo17) and each inbred line, $\beta$ is the fixed effect of that distance, $H$ is the tested locus, $\vartheta$ the effect of the locus, and ${\varepsilon }'$ is the vector of residuals assumed to be $\mathcal{N}(0,\sigma_{ {\varepsilon}'}^{2}I)$.  
SNPs were deemed to be statistically significant at $p\leq 0.001$. Analyses were also conducted controlling the false discovery rate \citep{Benjamini1995} at 10\%. 

\section*{Results and Discussion}

\subsection*{Prediction of deleterious mutations}

In order to investigate deleterious mutations in a diverse set of maize inbred lines, we first applied two complementary approaches to predict deleterious mutations across the maize genome.  We applied the software packages SIFT \citep{Ng2003, Ng2006} and MAPP \citep{Stone2005} to the 39,656 genes in version 5b of the maize filtered gene set  \citep[\url{http://www.maizesequence.org};][]{Schnable2009}. SIFT predicted amino acid change consequences for nearly 12 million codons in ~32,000 genes, while MAPP obtained predictions for a total of ~11 million codons in ~29,000 genes combined across the three ortholog datasets used (see methods). More than 80\% of predictions were congruent between the two approaches, similar to what has been seen in \emph{Arabidopsis thaliana} and rice \citep{Gunther2010}. SIFT and MAPP respectively identified $\sim$80\,\% and 60\, \% of amino acid polymorphisms as ``tolerated", with the remainder predicted to be premature stop codons or ``non-tolerated" amino acid changes; we will refer to these latter categories as predicted deleterious SNPs.

We then took advantage of recently published genotyping-by-sequencing \citep[GBS; ][]{Elshire2011} data to survey potentially deleterious mutations across a panel of 247 diverse maize inbred lines \citep{Larsson2013, Romay2013}. The genotyping data include a total of  437,650 SNPs which covered 112,326 and 107,472 codons representing 19,145 and 18,255 genes in the SIFT and MAPP data, respectively. Nearly 50\% of these codons showed no amino acid polymorphism in each dataset; while the vast majority of these monomorphic amino acids were due to synonymous polymorphisms in the GBS data, several hundred predicted deleterious amino acids were fixed across all maize lines analyzed (Supplemental Table~\ref{Predicitons}). Combining results from both SIFT and MAPP, our data consist of 25,352 predicted deleterious SNPs in 11,034 genes.

\subsection*{Characterization of deleterious SNPs in a diversity panel}

\begin{figure*}[!b]
  \begin{center}
   \includegraphics[width=150mm]{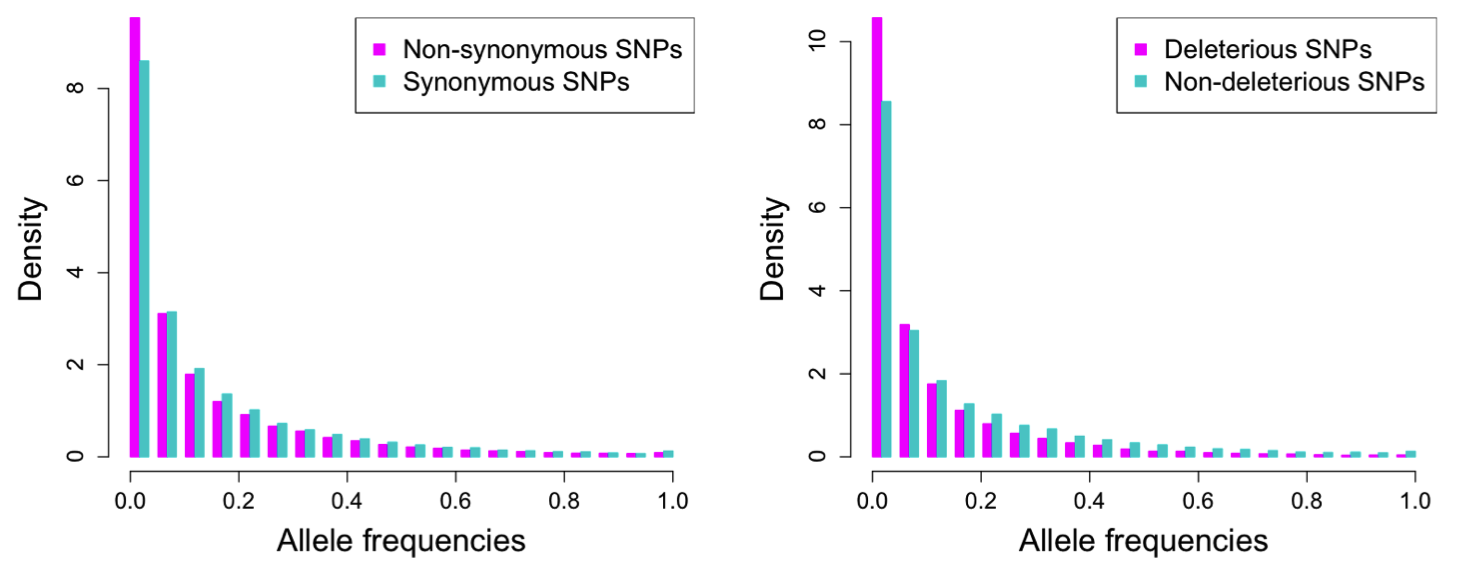}
    \caption{Derived site frequency spectrum of (A) synonymous \emph{vs} non-synonymous SNPs and (b) non-synonymous non-deleterious \emph{vs} non-synonymous deleterious SNPs. \emph{Tripsacum} was used as outgroup for identifying the derived allele.} 
\label{sfs_non_syn}
  \end{center}
\end{figure*}

Across all lines, the derived site frequency spectrum (SFS) of coding SNPs showed an excess of rare variants compared to neutral expectations, with 45\% of deleterious SNPs occurring at derived frequencies less than 5\% in the SFS across all lines. 
Even so, non-synonymous SNPs showed an excess of rare variants when compared to synonymous SNPs (Mann-Whitney U test p-value $<10^{-15}$; Figure~\ref{sfs_non_syn}-A), and predicted deleterious SNPs showed a marked excess of rare variants compared to other non-synonymous variants (Mann-Whitney U test p-value $<10^{-15}$; Figure~\ref{sfs_non_syn}-B). These observations are consistent with the action of weak purifying selection \citep{Cummings1998,Fay2001} and provide a measure of independent corroboration of the utility of MAPP and SIFT in predicting deleterious variants.

Although most predicted deleterious alleles were rare, 923 were found segregating at high frequency ($\geq 0.80$) across all lines.  To test whether these alleles may have been driven to high frequency by selection at linked loci during domestication \citep{Lu2006}, we analyzed the pattern of haplotype sharing across the genome \citep[PHS statistics;][]{Toomajian2006}. Only 87 of these SNPs (9.4\,\% of all tests) showed signs of positive selection in at least one of the genetic groups, and only 25 (2.7\,\%) were found in candidate regions for selection during maize domestication \citep{Hufford2012}, providing little evidence to support hitchhiking during domestication as a major influence on the distribution of deleterious alleles in the genome.

\begin{figure*}[!t]
  \begin{center}
   \includegraphics[width=140mm]{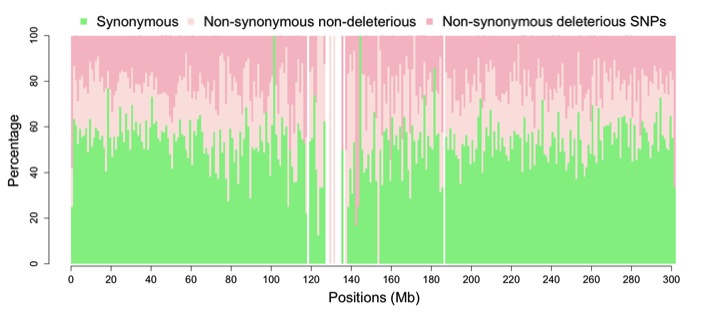}
    \caption{Proportion of genic SNPs predicted to be synonymous, non-synonymous non-deleterious and non-synonymous deleterious in 1\,Mb windows along chromosome\,1} 
   \label{non_syn_chr1}
  \end{center}
\end{figure*}

The proportion of genic SNPs predicted to be deleterious appeared relatively uniform (Figure \ref{non_syn_chr1} and  Supplemental Figure \ref{non_syn_chr1cM}) across the genome, with only a very low correlation  observed with recombination rate (Pearson's $r$ of 0.06; p-value =$0.005$).  Explicit comparison of 1,778 non-synonymous pericentromeric ($\pm$ 5 cM around the functional centromere) SNPs did not show an elevated proportion of predicted deleterious SNPs  in comparison to the whole genome (Fisher's Exact Test p-value $= 0.68$). 
The negative correlation between recombination and residual heterozygosity observed in recombinant inbred lines of the maize nested association mapping population has been attributed to the inefficiency of selection against deleterious alleles in low recombination regions of the genome \citep{McMullen2009,Gore2009}. 
Our results do not provide strong support for this explanation, perhaps suggesting that recombination in these regions over longer periods of time is sufficient to avoid the accumulation of deleterious alleles. Consistent with this idea, while regions of the \emph{Drosophila} genome completely lacking in recombination showed a severe reduction in the efficacy of selection, little difference was observed between regions with high and low rates of recombination \citep{Haddrill2007}.

Individual lines varied considerably in their content of predicted deleterious alleles, carrying between 4 and 16\% of all predicted deleterious alleles. Lines from the stiff stalk group carried on average fewer deleterious mutations (9\%) than did lines from other groups (14-15\%).  Although drift due to a  historically low $N_{e}$ \citep{Messmer1991} could explain this observation, other groups with low $N_{e}$ such as the popcorns do not show such a trend.  Instead, we posit that both the SIFT and MAPP algorithms may be biased against alleles found in the reference B73 genome which belongs to the stiff stalk heterotic group; similar bias has recently been described in analyses of the human genome \citep{Simons2013}.

\begin{figure*}[!t]
  \begin{center}
   \includegraphics[width=90mm]{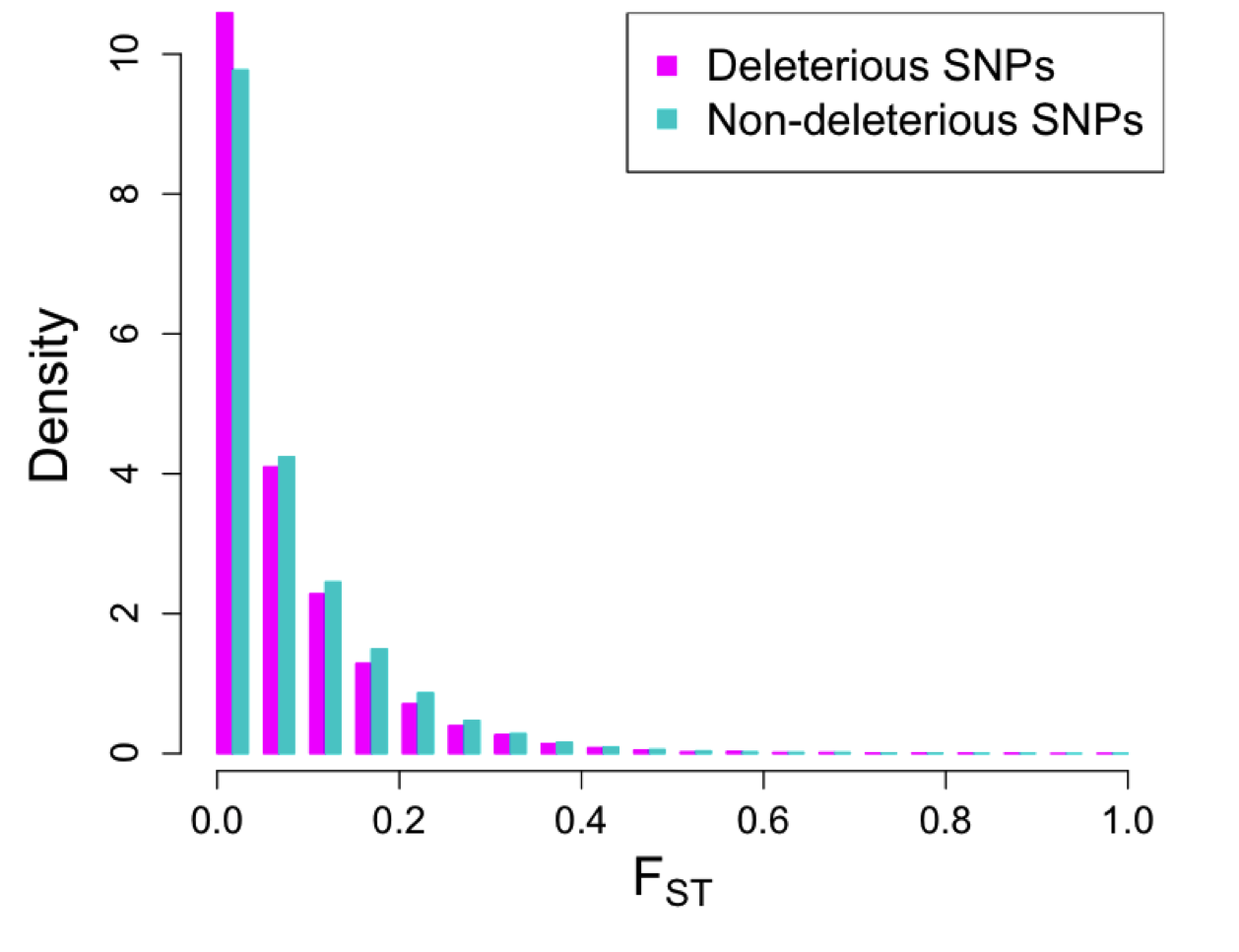}
    \caption{$\mathrm{F}_{\mathrm{ST}}$ distribution for deleterious and non-deleterious SNPs}
   \label{fst_dist}
  \end{center}
  \begin{center}
   \includegraphics[width=150mm]{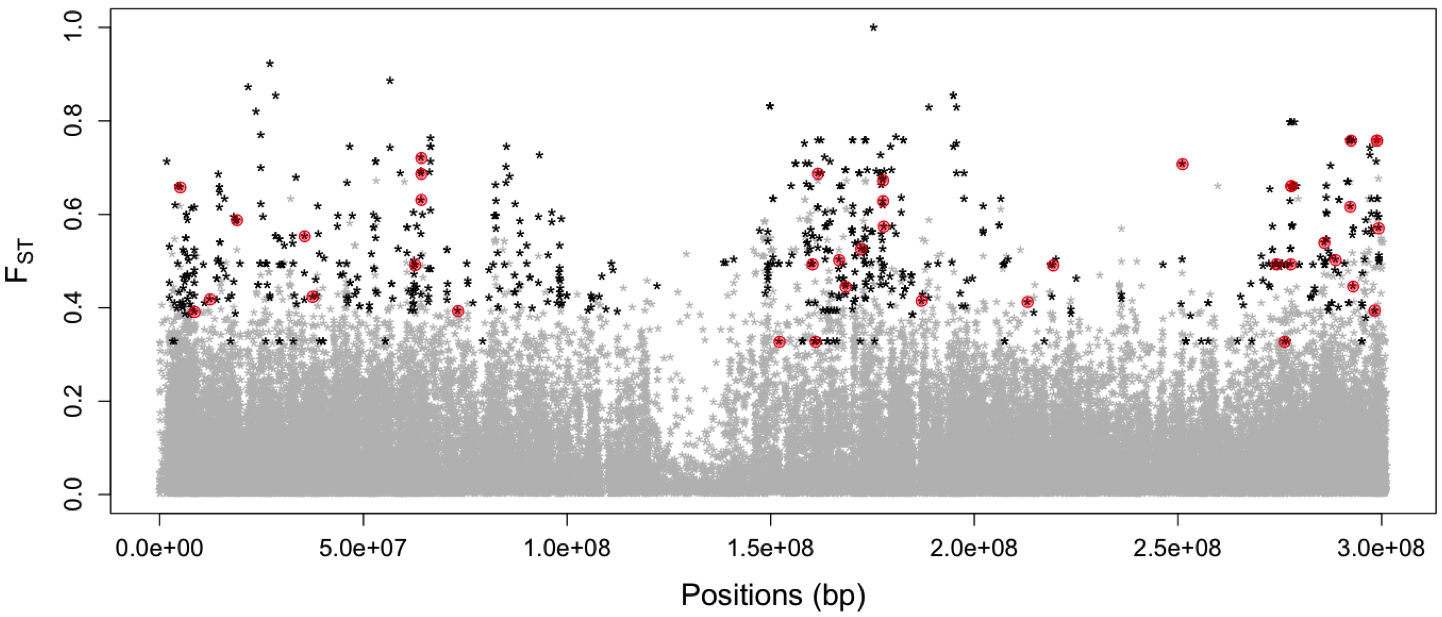}
    \caption{Distribution of $\mathrm{F}_{\mathrm{ST}}$ along chromosome 1; black dots represent top 1\% SNPs, outlier predicted deleterious SNPs are surrounded in red.} 
   \label{fst_chr1}
  \end{center}
\end{figure*}

Allele sharing at predicted deleterious SNPs generally followed genome-wide patterns of identity by state (IBS). Within the non-stiff stalk, tropical, popcorn and sweet groups, correlations were generally high (Pearson's $r$ of 0.75-0.99) between numbers of shared predicted deleterious alleles (mean of 5 -10\%) and IBS. Correlations between inbreds from different genetic groups were much lower ($r$ of 25 - 52 \%), however, as has been previously seen in correlations  between IBS and heterosis observed at SSR loci \citep{Flint-Garcia2009}. The  ``mixed'' (within group $r=0.22$ and $r=-0.05\ to\ 0.36$ with other groups) and stiff stalk (within-group $r=0.15$ and $r=-0.65\ to\ 0.16$ with other groups) groups appeared exceptions to this pattern, perhaps due to the aforementioned ascertainment bias or previously unrecognized population substructure within these groups(Supplemental Figure~\ref{figureS3}). 

Across all genetic groups, levels of population differentiation were slightly lower for predicted deleterious (mean $\mathrm{F}_{\mathrm{ST}}=0.07$) than non-deleterious (mean $\mathrm{F}_{\mathrm{ST}}=0.08$) SNPs (Mann-Whitney U test p-value $< 10^{-15}$ ; Figure~\ref{fst_dist}). After correcting 
for allele frequencies in both classes, however, these differences disappeared, and the proportion of deleterious SNPs in the top 1\% of $\mathrm{F}_{\mathrm{ST}}$ was not significantly different from the proportion observed for synonymous SNPs (Fisher's Exact Test p-value $= 0.94$) or all SNPs in genic regions (Fisher's Exact Test p-value $= 0.51$).
After allele frequency correction, 287 genes had a predicted deleterious SNP in the top 1\% of Fst among genetic groups, and 30 genes had 2 or more high-Fst predicted deleterious SNPs.  Only eleven genes ($4\%$) with high-Fst deleterious SNPs  are, however, found in regions thought to be selected during maize improvement  \citep{Hufford2012}, and only 44 genes ($15\%$) show significant signs of positive selection with the PHS statistic, again providing little support for the idea that these SNPs were affected by selection on linked beneficial mutations.

\begin{figure*}[!b]
  \begin{center}
   \includegraphics[width=110mm]{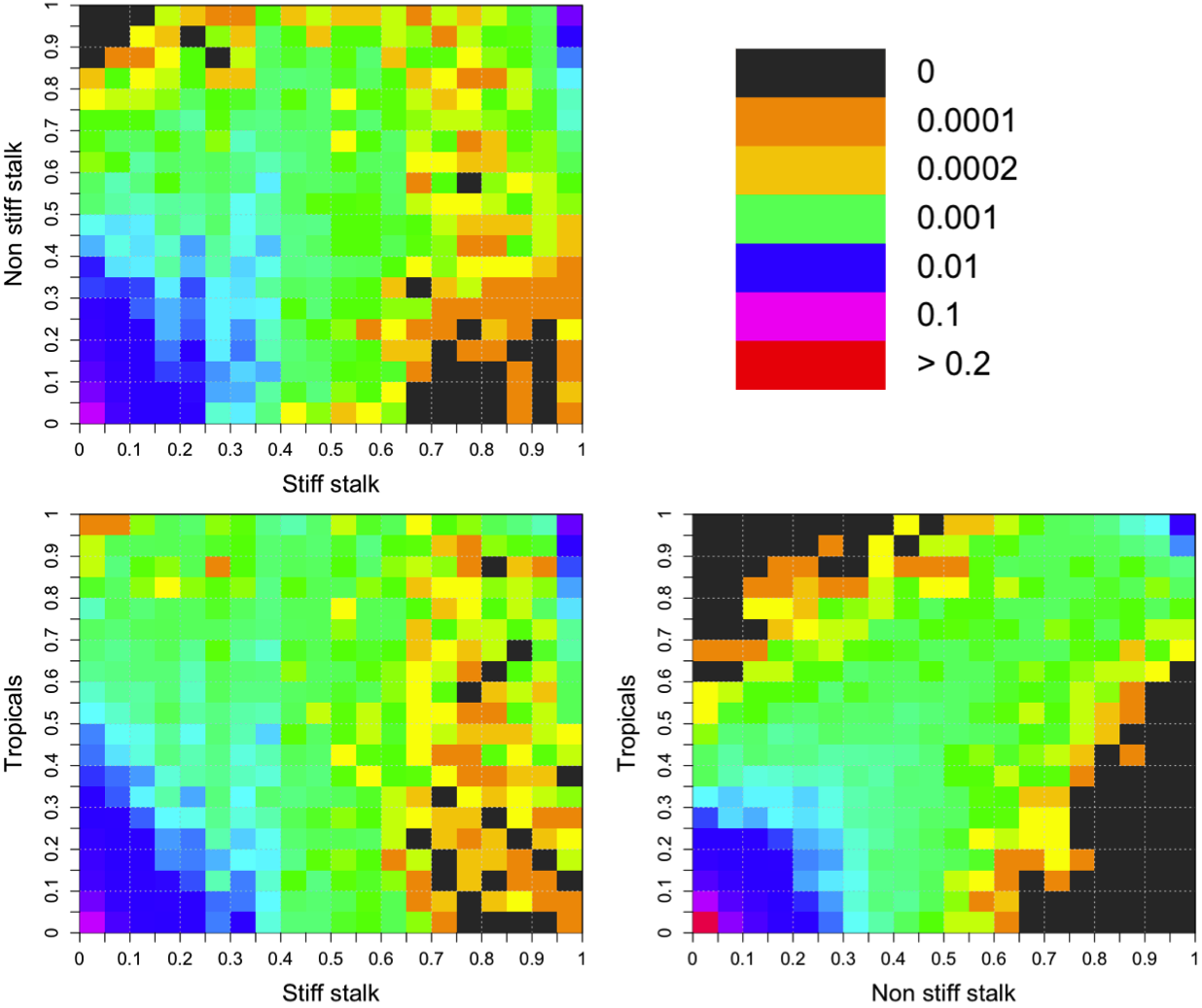}
    \caption{Joint site frequency spectrum of stiff-stalk, non stiff-stalk and tropical inbred lines. Shown is the frequency of the predicted deleterious allele.}
   \label{jfs}
  \end{center}
\end{figure*}

Comparisons of the predicted deleterious SFS between stiff stalk, non stiff stalk, and tropical groups (Figure~\ref{jfs}) mirrored patterns of between-group  $\mathrm{F}_{\mathrm{ST}}$, revealing few fixed differences between groups and generally low frequencies within groups, as well as higher differentiation in comparisons involving the stiff stalk group (Figure~\ref{jfs}).

\subsection*{Effect of deleterious mutations on traits of interest}

To investigate the contribution of predicted deleterious alleles to observed levels of heterosis and inbreeding depression, we performed a genome wide association analysis of 17 traits evaluated in two populations (see Methods). Analyses were carried out using the genetic values of inbred lines and both mid-parent and best-parent heterosis. Genome wide association results using the genetic values of inbred lines identified between 219 (cob diameter) and 598 (cob length) significant SNPs with a high proportion of genic loci (up to 70\%) but little evidence for significant enrichment of predicted deleterious SNPs (Table \ref{popA_sig_enrichment} and Supplemental Table \ref{popB_sig_enrichment}). 

\begin{table*}[!t]
  \begin{center}

    \caption[]{Total number of significant SNPs (\emph{n}) and fold enrichment (\emph{f}) in genic regions, for loci with deleterious mutations in population A. Numbers marked with ``*'' are statistically significant.}
{\fontsize{10}{10}\sf
      \begin{tabular}{l|rr|rr|rr} 
\multicolumn{1}{c}{}	&	\multicolumn{2}{c}{Inbreds}	&	\multicolumn{2}{c}{BPH}	&	\multicolumn{2}{c}{MPH}	\\	\hline 
Traits	& 	\multicolumn{1}{c}{\emph{n}} 	& 	\multicolumn{1}{c|}{\emph{f}}	& \multicolumn{1}{c}{\emph{n}} 	& 	\multicolumn{1}{c}{\emph{f}} 		& \multicolumn{1}{c}{\emph{n}} 	& 	\multicolumn{1}{c}{\emph{f}} 		\\	\hline \hline  
DTT 	& 	475 	& 	1.05 	& 	3372 	& 	1.15* 	& 	1123 	& 	1.12	\\
TSLLEN 	& 	458 	& 	0.81 	& 	297 	& 	1.21 	& 	365 	& 	1.16	\\
TSLBCHCNT 	& 	300 	& 	0.98 	& 	4077 	& 	0.98 	& 	1257 	& 	1.12	\\
TSLANG 	& 	244 	& 	1.11 	& 	490 	& 	0.93 	& 	646 	& 	1.18	\\
PLTHT 	& 	282 	& 	0.92 	& 	18068 	& 	0.98 	& 	9712 	& 	0.93	\\
UPLFANG 	& 	415 	& 	1.20 	& 	8927 	& 	0.99 	& 	2266 	& 	1.12	\\
LFWDT 	& 	289 	& 	1.21 	& 	1064 	& 	1.16 	& 	1051 	& 	1.01	\\
LFLEN 	& 	389 	& 	1.14 	& 	4256 	& 	0.93 	& 	2257 	& 	1.07	\\
KNLHGT 	& 	292 	& 	1.10 	& 	8752 	& 	1.08 	& 	4512 	& 	1.01	\\
RPR 	& 	258 	& 	0.79 	& 	359 	& 	1.30 	& 	375 	& 	0.93	\\
PLTYLD 	& 	257 	& 	1.50 	& 	7440 	& 	1.12* 	& 	7007 	& 	1.14*	\\
EARLGH 	& 	231 	& 	0.89 	& 	605 	& 	1.11* 	& 	907 	& 	1.00	\\	
10KWT 	& 	298 	& 	1.29 	& 	709 	& 	1.15 	& 	761 	& 	1.30	\\
COBDIA 	& 	219 	& 	1.04 	& 	4363 	& 	1.16* 	& 	405 	& 	0.88	\\
COBWT 	& 	228 	& 	1.09 	& 	1746 	& 	0.93 	& 	519 	& 	0.69	\\
TOTKNLWT 	& 	256 	& 	0.88 	& 	3781 	& 	0.98 	& 	2045 	& 	0.95	\\
      \end{tabular}
}
    \label{popA_sig_enrichment}  
  \end{center}
\end{table*}

Observed frequencies of deleterious SNPs in different populations (Figure \ref{jfs}) may help explain patterns of hybrid vigor.  Though $\mathrm{F}_{\mathrm{ST}}$ is generally low, inbreds from different genetic groups are nonetheless likely to share fewer deleterious variants than inbreds from the same group, and heterosis is higher among crosses between groups (Supplemental Figure~\ref{figureS4}).  Nonetheless, even crosses among inbreds from the same genetic group show evidence of heterosis (Supplemental Figure~\ref{figureS1}-A), likely due to the large number of deleterious SNPs segregating at low frequencies within individual populations.   

Results for association between SNP heterozygosity and heterosis showed highly variable numbers of significant loci (Table \ref{popA_sig_enrichment} and Supplemental Table \ref{popB_sig_enrichment}), also with a high proportion of genic SNPs (up to 74\%). Significant SNPs explained between 4 and 40\% of the observed phenotypic variation in heterosis; though these values are likely inflated due to small sample size \citep{Beavis1994}. The highest numbers of associated loci were observed for plant height and yield-related traits which also showed the highest levels of observed heterosis. 
Furthermore, most traits exhibited some enrichment ($5-45\%$) of predicted deleterious SNPs and the enrichment was statistically significant for whole plant yield and days to tasseling.

Because most deleterious SNPs are at frequencies too low for inclusion in association analyses (Figure~\ref{sfs_non_syn}), we expanded our test of enrichment to the gene level, asking whether genes with predicted deleterious SNPs were more likely than random to have SNPs significantly associated with traits of interest. At this level we see much stronger evidence of enrichment; a number of traits show statistically significant enrichment in population\,A, but virtually all traits in both populations show a positive enrichment for genes with predicted deleterious SNPs (Tables \ref{gene_enrichment_A} and Supplemental Table \ref{gene_enrichment_B}), a result that is highly unlikely by chance (sign test p-value=$3\ 10^{-5}$ for population\,A and $0.01$ for population\,B). Similar tests of low-frequency synonymous SNPs show no evidence of enrichment (p-value $\approx 1$), and the low correlation between total SNPs in a gene and the number of significant associations ($r\leq 0.2$) suggests that our observation is not an artifact of the number of SNPs analyzed per gene. 

We posit that the observed excess of significant associations in genes with predicted deleterious variants may be due to so-called synthetic associations between rare deleterious loci and a common locus at high enough frequency to be included in association mapping tests \citep{Dickson2010,Goldstein2009}. Recent work suggests that this sort of association is only likely to hold for deleterious SNPs with a relatively small effect on phenotype \citep{Thornton2013}, which is consistent with the expected weak to intermediate effects of  deleterious loci likely to be involved in heterosis \citep{Charlesworth1987,Whitlock2000,Glemin2003,Charlesworth2009}. Strongly deleterious SNPs, though potentially playing a role in inbreeding depression \citep{Whitlock2000}, are less likely to be observed in our study as selection should effectively remove them from our panel of inbred lines.  

\begin{table*}[!t]
  \begin{center}
    \caption[]{Total number of genes with significant SNPs (\emph{n}) and fold enrichment for genes with predicted deleterious SNPs(\emph{f}) in population A}
{\fontsize{10}{10}\sf
      \begin{tabular}{l|rr|rr|rr} 
\multicolumn{1}{c}{}	&	\multicolumn{2}{c}{Inbreds}	&	\multicolumn{2}{c}{BPH}	&	\multicolumn{2}{c}{MPH}	\\	\hline 
Traits	& 	\multicolumn{1}{c}{\emph{n}} 	& 	\multicolumn{1}{c|}{\emph{f}}	& \multicolumn{1}{c}{\emph{n}} 	& 	\multicolumn{1}{c}{\emph{f}} 		& \multicolumn{1}{c}{\emph{n}} 	& 	\multicolumn{1}{c}{\emph{f}} 		\\	\hline \hline 
DTT	&	176	&	1.11	&	1137	&	1.12*	&	429	&	1.15*		\\
TSLLEN	&	173	&	1.08	&	128	&	1.14	&	154	&	1.20	\\
TSLBCHCNT	&	114	&	1.02	&	1257	&	1.13*	&	472	&	1.14*		\\
TSLANG	&	103	&	1.03	&	177	&	1.10	&	254	&	1.15	\\
PLTHT	&	128	&	1.22	&	4529	&	1.10*	&	2741	&	1.10*	\\
UPLFANG	&	166	&	1.13	&	2553	&	1.11*	&	810	&	1.15*	\\
LFWDT	&	112	&	1.27	&	379	&	1.05	&	375	&	1.14	\\
LFLEN	&	141	&	1.18	&	1290	&	1.13*	&	821	&	1.20*	\\
KNLHGT	&	123	&	1.09	&	2633	&	1.13*	&	1506	&	1.14	\\
RPR	&	99	&	1.24	&	150	&	1.15	&	145	&	1.07	\\
PLTYLD	&	117	&	1.22	&	2440	&	1.14*	&	2302	&	1.14*		\\
EARLGH	&	84	&	1.02	&	230	&	1.20	&	333	&	1.15	\\
10KWT	&	137	&	1.18	&	288	&	1.17	&	308	&	1.13	\\
COBDIA	&	90	&	1.10	&	1419	&	1.13*	&	162	&	1.12	\\
COBWT	&	99	&	1.19	&	548	&	1.07	&	176	&	1.13	\\
TOTKNLWT	&	101	&	1.18	&	1228	&	1.11*	&	714	&	1.07	\\
      \end{tabular}
}
    \label{gene_enrichment_A}  
  \end{center}
\end{table*}

Although we have analyzed only a relatively small subset of the genome-wide diversity of maize \citep{Chia2012}, our data nonetheless present the first genome-wide scan of deleterious coding variants in maize.  Our results provide evidence for the contribution of deleterious mutations to heterosis via complementation, consistent with the dominance hypothesis. 
The weak expected effects of these deleterious loci, combined with their low frequencies, make their detection difficult using the conventional approaches. \emph{A priori} prediction of the potential effect of rare polymorphisms, however, may improve predictions of inbred line breeding values and combining ability. Future analysis of full genome sequence data, allowing for the inclusion of all coding SNPs and noncoding variants, will provide an even richer catalog of variants that will expand our understanding of the role of rare deleterious variants in maize breeding.

\subsubsection*{Acknowledgments}
We would like to thank S. Flint-Garcia and S. Takuno for help with data analysis, E.S. Buckler for early access to the genotyping data, and G. Coop, J. Gerke, P. Morrell, P. Ralph, and O. Smith for helpful comments. 
This project was supported by Agriculture and Food Research Initiative Competitive Grant 2009-01864 from the USDA National Institute of Food and Agriculture as well as a grant from DuPont Pioneer. 

\clearpage
\bibliography{Heterosis}

\clearpage
\setcounter{figure}{0}
\setcounter{table}{0}
\renewcommand{\figurename}{Sup. Fig.}
\renewcommand{\tablename}{Sup. Table}

\section*{Supplementals}

\subsection*{List of the inbred lines used } 
\subsubsection*{ PopulationA}
B73, A214N, A441.5, A554, A556, A6, A619, A632, A634, A635, A641, A654, A659, A661, A679, A680, A682, AB28A, B10, B104, B105, B109, B115, B14A, B164, B2, B37, B46, B57, B64, B68, B73HTRHM, B75, B76, B77, B79, B84, B97, CH701.30, CH9, CI187.2, CI21E, CI28A, CI31A, CI3A, CI64, CI66, CI7, CI90C, CI91B, CM174, CM37, CM7, CML10, CML103, CML108, CML11, CML14, CML154Q, CML157Q, CML158Q, CML218, CML220, CML228, CML238, CML247, CML258, CML261, CML264, CML277, CML281, CML287, CML311, CML314, CML321, CML322, CML323, CML328, CML331, CML332, CML333, CML341, CML38, CML5, CML52, CML69, CML77, CML91, CML92, CMV3, CO255, D940Y, DE1, DE2, DE811, E2558W, EP1, F2834T, F44, F6, GA209, GT112, H105W, H84, H91, H95, H99, HI27, HP301, HY, I137TN, I205, I29, IA2132, IA5125, IDS28, IDS69, IDS91, IL101T, IL14H, IL677A, K148, K4, K55, K64, KI11, KI14, KI2021, KI21, KI3, KI43, KI44, KY21, KY226, KY228, L317, L578, M14, M162W, M37W, MEF156.55.2, MO17, MO18W, MO1W, MO24W, MO44, MO45, MO46, MOG, MP339, MS1334, MS153, MS71, MT42, N192, N28HT, N6, N7A, NC222, NC230, NC232, NC236, NC238, NC250, NC258, NC260, NC262, NC264, NC294, NC296, NC296A, NC298, NC300, NC302, NC304, NC306, NC310, NC314, NC318, NC320, NC324, NC326, NC328, NC33, NC336, NC338, NC342, NC344, NC346, NC348, NC350, NC352, NC354, NC356, NC358, NC360, NC362, NC364, NC366, NC368, ND246, OH40B, OH43E, OH603, OH7B, OS420, P39, PA762, PA875, PA880, PA91, R168, R177, R229, R4, SA24, SC357, SC55, SD44, SG1533, SG18, T232, T8, TX303, TZI10, TZI11, TZI16, TZI18, TZI25, TZI8, TZI9, U267Y, VA102, VA14, VA22, VA35, VA59, VA99, VAW6, W117HT, W153R, W182B, W64A, WD, X33.16, X38.11, X4226, X4722

\subsubsection*{ PopulationB}
B73, MO17, X33.16, A188, A239, A619, A632, A634, A635, A641, A654, A661, A679, A680, A682, B103, B104, B109, B115, B14A, B37, B46, B52, B57, B64, B68, B73, B73HTRHM, B75, B76, B77, B79, B84, C103, C49A, CH701.30, CM105, CM174, CO125, DE.2, DE1, DE811, EP1, H105W, H49, H84, H91, H95, H99, HP301, IL101, IL14H, K148, KY226, M14, MEF156.55.2, MO44, MO45, MO46, MO47, MS1334, MS153, MS71, N192, N28HT, N6, NC262, NC264, NC294, NC306, NC310, NC314, NC324, NC326, NC328, NC342, NC364, ND246, OH43, OH43E, OS420, P39, PA762, PA875, PA880, PA91, R168, R177, R4, SD40, SD44, SG18, VA102, VA14, VA17, VA22, VA35, VA85, VA99, W182B, W22, W64A, WF9, YU796.NS.

\begin{figure*}[h]
  \begin{center}
   \includegraphics[width=150mm]{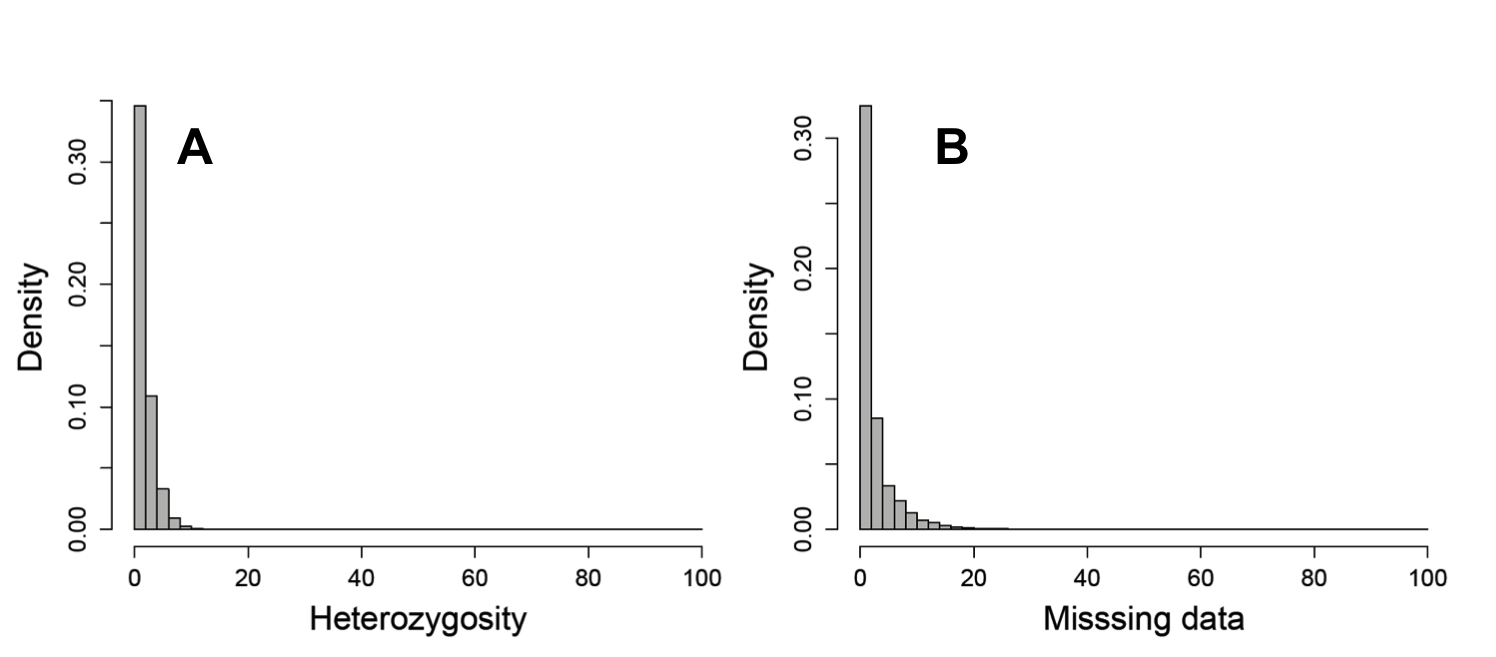}
    \caption{Histograms of the percentage of (A) heterozygosity and (B) missing data per SNP}
   \label{figureS1}
  \end{center}
\end{figure*}

\subsection*{List of genomes used for reciprocal BLAST}
\it{Aquilegia coerulea}, 
\it{Arabidopsis lyrata}, 
\it{Arabidopsis thaliana}, 
\it{Brachypodium distachyon}, 
\it{Brassica rapa}, 
\it{Capsella rubella}, 
\it{Carica papaya}, 
\it{Chlamydomonas reinhardtii}, 
\it{Citrus clementina}, 
\it{Citrus sinensis}, 
\it{Cucumis sativus}, 
\it{Eucalyptus grandis}, 
\it{Glycine max}, 
\it{Linum usitatissimum}, 
\it{Malus domestica}, 
\it{Manihot esculenta}, 
\it{Medicago truncatula}, 
\it{Mimulus guttatus}, 
\it{Oryza sativa}, 
\it{Panicum virgatum}, 
\it{Phaseolus vulgaris}, 
\it{Physcomitrella patens}, 
\it{Populus trichocarpa}, 
\it{Prunus persica}, 
\it{Ricinus communis}, 
\it{Selaginella moellendorffii}, 
\it{Setaria italica}, 
\it{Sorghum bicolor}, 
\it{Thellunigiella halophila}, 
\it{Vitis vinifera}, 
\it{Volvox carteri}.

\begin{table*}[ht]
  \begin{center}
{\fontsize{10}{10}\sf
\caption[]{List of Analyzed traits}
\begin{tabular}{llc} \hline
Traits & Abbreviation & Populations \\	 \hline\hline
Days to tasseling	&	DTT	&	A	\\
Tassel length (cm)	&	TSLLEN	&	A	\\
Tassel branch count	&	TSLBCHCNT	&	A	\\
Tassel angle	&	TSANG	&	A	\\
Plant height (cm)	&	PLTHT	&	A \& B	\\
Upper leaf angle	&	UPLFANG	&	A	\\
Leaf width (cm)	&	LFWDT	&	A	\\
Leaf length (cm)	&	LFLEN	&	A	\\
Kernel heigth	&	KNLHGT	&	A	\\
Stem puncture resistance (kg/section)	&	RPR	&	A	\\
Plant yield (g/plant)	&	PLTYLD	&	A	\\
Ear length (cm)	&	EARLGH	&	A \& B	\\
10 kernel weight (g)	&	10KWT	&	A	\\
Cob diameter (cm)	&	COBDIA	&	A \& B	\\
Cob weight (g)	&	COBWT	&	A \& B	\\
Seed number per ear	&	SEEDNB	&	B	\\
\end{tabular}
\label{Traits}  
}
  \end{center}
\end{table*}

\begin{table*}[ht]
{\fontsize{10}{10}\sf
\caption[]{Detailed results of the prediction of deleterious amino acids with MAPP, using the different gene sets, and with SIFT}
\begin{tabular}{lrrrr} \hline
\multicolumn{1}{c}{}	&	\multicolumn{3}{c}{MAPP}	&	\multicolumn{1}{c}{SIFT}	\\	\hline
Gene sets  & BLASTX & Reciprocal BLAST & Syntenic genes	&	PSI-BLAST \\	 \hline\hline
Total a.a. positions with predictions 	& 7,746,638 &	5,570,035 &	6,869,010	&	11,906,167 \\
Total number of genes &	20,348 &	11,918 &	17,957	&	31,843	\\
Number of positions covered by SNPs &	74,909 &	52,283 &	72,562	&	112,326	\\
Number of genes covered by SNPs &	12,561	&	8,553	&	12,615	&	19,145	\\
Monomorphic tolerated &	39,009 &	25,270 &	39,300	&	58,685	\\
Monomorphic not tolerated* &	144 &	3470 &	14	&	387	\\
Polymorphic tolerated	& 18,379 &	10,753 &		17,792	&	42,606	\\
Polymorphic not tolerated* &	 17,377 &	12,790 &	15456	&	10,648	\\
\multicolumn{5}{l}{*Includes premature stop codons}
\end{tabular}
\label{Predicitons} 
}
\end{table*}

\begin{table*}[ht]
  \begin{center}
    \caption[]{Comparion of the results of MAPP predictions with the different gene sets.} 
{\fontsize{10}{10}\sf
      \begin{tabular}{lccc}
{Gene sets}		&	{      BLASTX      }	&	{      Reciprocal BLAST      }	&	{Syntenic genes}	\\ \hline \hline
BLASTX	&	-	&	80.1\%	&	78.2\%	\\
Reciprocal BLAST	&	38,054 (6,169)	&	-	&	79.8\%	\\
Syntenic genes		&	45,412 (7,745)	&	32,222 (5,488)	&	-	\\
\multicolumn{4}{l}{The lower triangle indicates the number of  amino acid positions predicted with two given }	\\
\multicolumn{4}{l}{gene sets and covered by GBS SNPs (number of genes between brackets);  the upper }	\\
\multicolumn{4}{l}{triangle indicates the percentage of amino acids with the same predictions.}\\
      \end{tabular}
    \label{tableS2} 
}
  \end{center}
\end{table*}

\begin{table*}[ht]
  \begin{center}
    \caption[]{Total number of significant SNPs (\emph{n}) and fold enrichment (\emph{f}), in genic regions, for loci with deleterious mutations in population B. Numbers marked with * are statistically significant.}
{\fontsize{10}{10}\sf
      \begin{tabular}{l|rr|rr|rr|rr|rr} 
\multicolumn{1}{c}{}	&	\multicolumn{2}{c}{Inbreds}	&	\multicolumn{2}{c}{BPH\_B73}	&	\multicolumn{2}{c}{MPH\_B73}	&	\multicolumn{2}{c}{BPH\_Mo17}	&	\multicolumn{2}{c}{MPH\_Mo17}	\\	\hline 
Traits	& 	\multicolumn{1}{c}{\emph{n}} 	& 	\multicolumn{1}{c|}{\emph{f}} 	& 	\multicolumn{1}{c}{\emph{n}} 	& 	\multicolumn{1}{c|}{\emph{f}} 	& 	\multicolumn{1}{c}{\emph{n}} 	& 	\multicolumn{1}{c|}{\emph{f}}	& 	\multicolumn{1}{c}{\emph{n}} 	& 	\multicolumn{1}{c|}{\emph{f}} 	& 	\multicolumn{1}{c}{\emph{n}} 	& 	\multicolumn{1}{c}{\emph{f}}	\\	\hline \hline 
10KWT 	& 	310 	& 	0.77 	& 	404 	& 	1.17* 	& 	257 	& 	0.86 	& 	698 	& 	0.83 	& 	723 	& 	0.98	\\
COBWT 	& 	313 	& 	0.62 	& 	941 	& 	1.15* 	& 	387 	& 	0.69 	& 	257 	& 	1.33 	& 	532 	& 	0.95	\\
COBDIA 	& 	226 	& 	1.49 	& 	159 	& 	1.25* 	& 	236 	& 	1.06* 	& 	349 	& 	0.78 	& 	615 	& 	0.72	\\
COBLEN 	& 	598 	& 	1.08 	& 	239 	& 	1.20* 	& 	97 	& 	0.24 	& 	280 	& 	1.08 	& 	140 	& 	0.92	\\
SEEDWT 	& 	362 	& 	1.09 	& 	378 	& 	1.32* 	& 	118 	& 	1.23* 	& 	1043 	& 	0.92 	& 	1080 	& 	0.78	\\
SEEDNB 	& 	373 	& 	0.99 	& 	320 	& 	0.86 	& 	251 	& 	0.92 	& 	348 	& 	1.06 	& 	454 	& 	0.82	\\
PLTHT 	& 	505 	& 	1.02 	& 	261 	& 	0.89 	& 	143 	& 	1.45* 	& 	1022 	& 	1.08 	& 	156 	& 	1.16	\\
      \end{tabular}
}
    \label{popB_sig_enrichment}  
  \end{center}
\end{table*}

\begin{table*}[ht]
  \begin{center}
    \caption[]{Total number of genes with significant SNPs (\emph{n}) and fold enrichment for genes with predicted deleterious SNPs (\emph{f}) in population B}
{\fontsize{10}{10}\sf
      \begin{tabular}{l|rr|rr|rr|rr|rr} 
\multicolumn{1}{c}{}	&	\multicolumn{2}{c}{Inbreds}	&	\multicolumn{2}{c}{BPH\_B73}	&	\multicolumn{2}{c}{MPH\_B73}	&	\multicolumn{2}{c}{BPH\_Mo17}	&	\multicolumn{2}{c}{MPH\_Mo17}	\\	\hline 
Traits	& 	 \multicolumn{1}{c}{\emph{n}} 	& 	\multicolumn{1}{c|}{\emph{f}} 	& 	\multicolumn{1}{c}{\emph{n}} 	& 	\multicolumn{1}{c|}{\emph{f}} 	& 	\multicolumn{1}{c}{\emph{n}} 	& 	\multicolumn{1}{c|}{\emph{f}}	& 	\multicolumn{1}{c}{\emph{n}} 	& 	\multicolumn{1}{c|}{\emph{f}}	& 	\multicolumn{1}{c}{\emph{n}}	&	\multicolumn{1}{c}{\emph{f}}	\\	\hline \hline 
10KWT	&	73	&	1.17	&	169	&	1.14	&	95	&	1.11	&	246	&	1.11	&	274	&	1.11	\\
COBWT	&	71	&	1.13	&	316	&	1.08	&	128	&	1.04	&	94	&	1.10	&	204	&	1.10	\\
COBDIA	&	81	&	1.07	&	57	&	1.08	&	86	&	1.11	&	134	&	1.03	&	234	&	1.14	\\
COBLEN	&	203	&	1.09	&	89	&	1.24	&	30	&	1.17	&	110	&	1.17	&	51	&	1.21	\\
SEEDWT	&	138	&	1.10	&	146	&	1.14	&	50	&	0.97	&	371	&	1.09	&	389	&	1.09	\\
SEEDNB	&	106	&	1.15	&	128	&	1.13	&	116	&	0.98	&	130	&	1.12	&	166	&	1.09	\\
PLTHT	&	169	&	1.15	&	112	&	1.09	&	65	&	1.13	&	348	&	1.15	&	65	&	1.15	\\
      \end{tabular}
}
    \label{gene_enrichment_B}  
  \end{center}
\end{table*}

\begin{figure*}[h]
  \begin{center}
   \includegraphics[width=150mm]{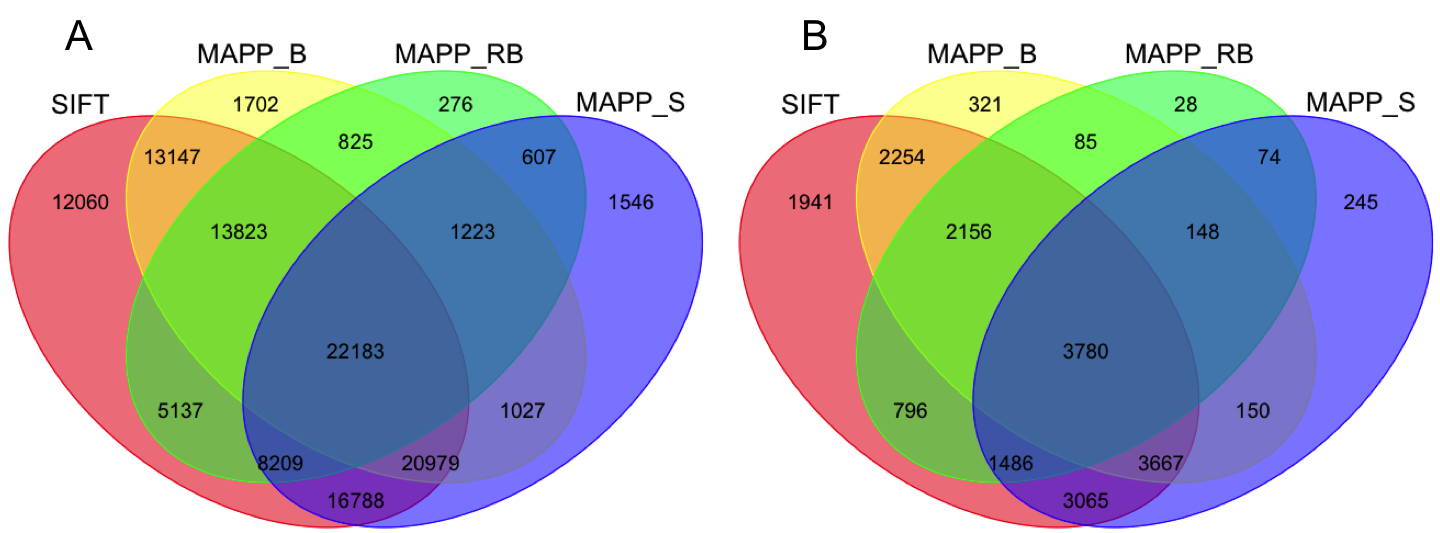}
    \caption{Comparison of the number of predicted (A) amino acids and (B) genes, covered by SNP data. For MAPP, 3 gene sets were used: BLASTX (MAPP\_B), reciprocal BLAST (MAPP\_RB) and syntenic genes (MAPP\_S)}
   \label{figureS2}
  \end{center}
\end{figure*}

\begin{figure*}[ht]
  \begin{center}
   \includegraphics[width=140mm]{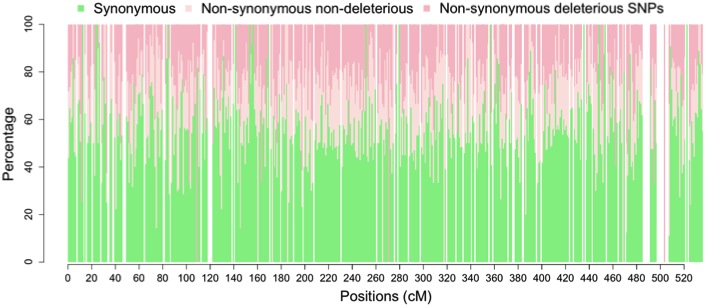}
    \caption{Proportion of genic SNPs predicted to be synonymous, non-synonymous non-deleterious and non-synonymous deleterious in 1\,cM windows along chromosome\,1} 
   \label{non_syn_chr1cM}
  \end{center}
\end{figure*}

\begin{figure*}[h]
  \begin{center}
   \includegraphics[width=150mm]{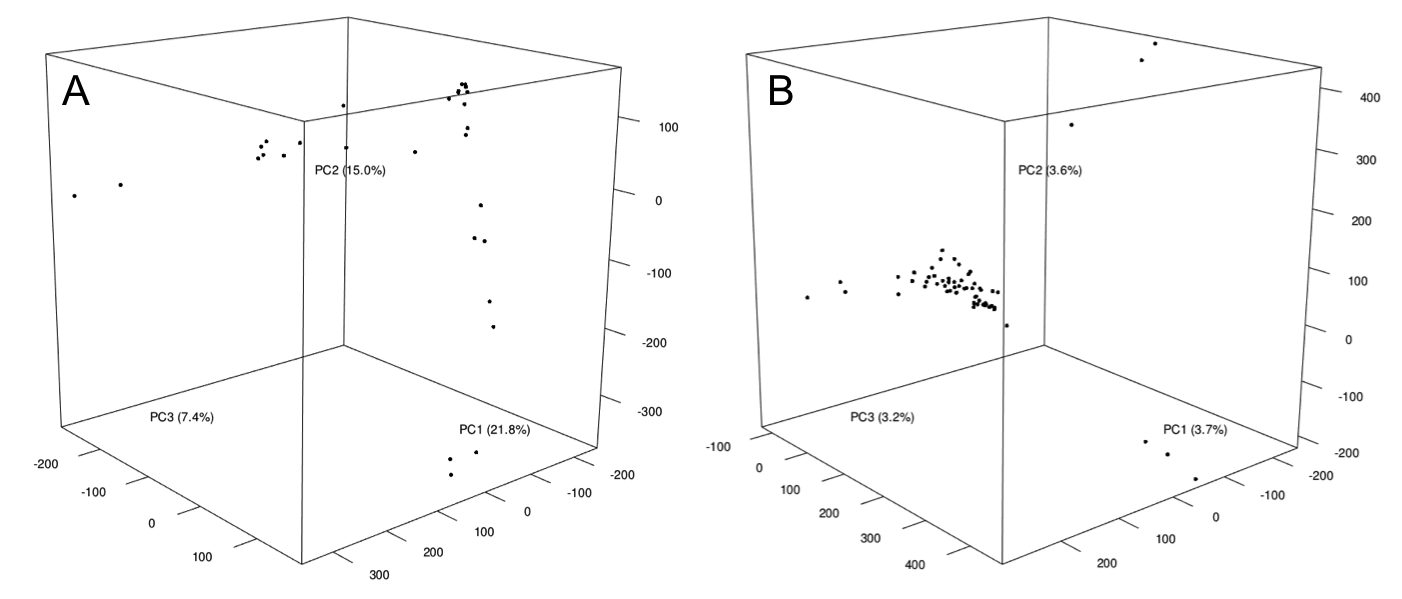}
    \caption{Projection of the (A) stiff stalk and (B) mixed inbred lines on the three first axes of a principal component analysis}
   \label{figureS3}
  \end{center}
\end{figure*}


\begin{figure*}[h]
  \begin{center}
   \includegraphics[width=100mm]{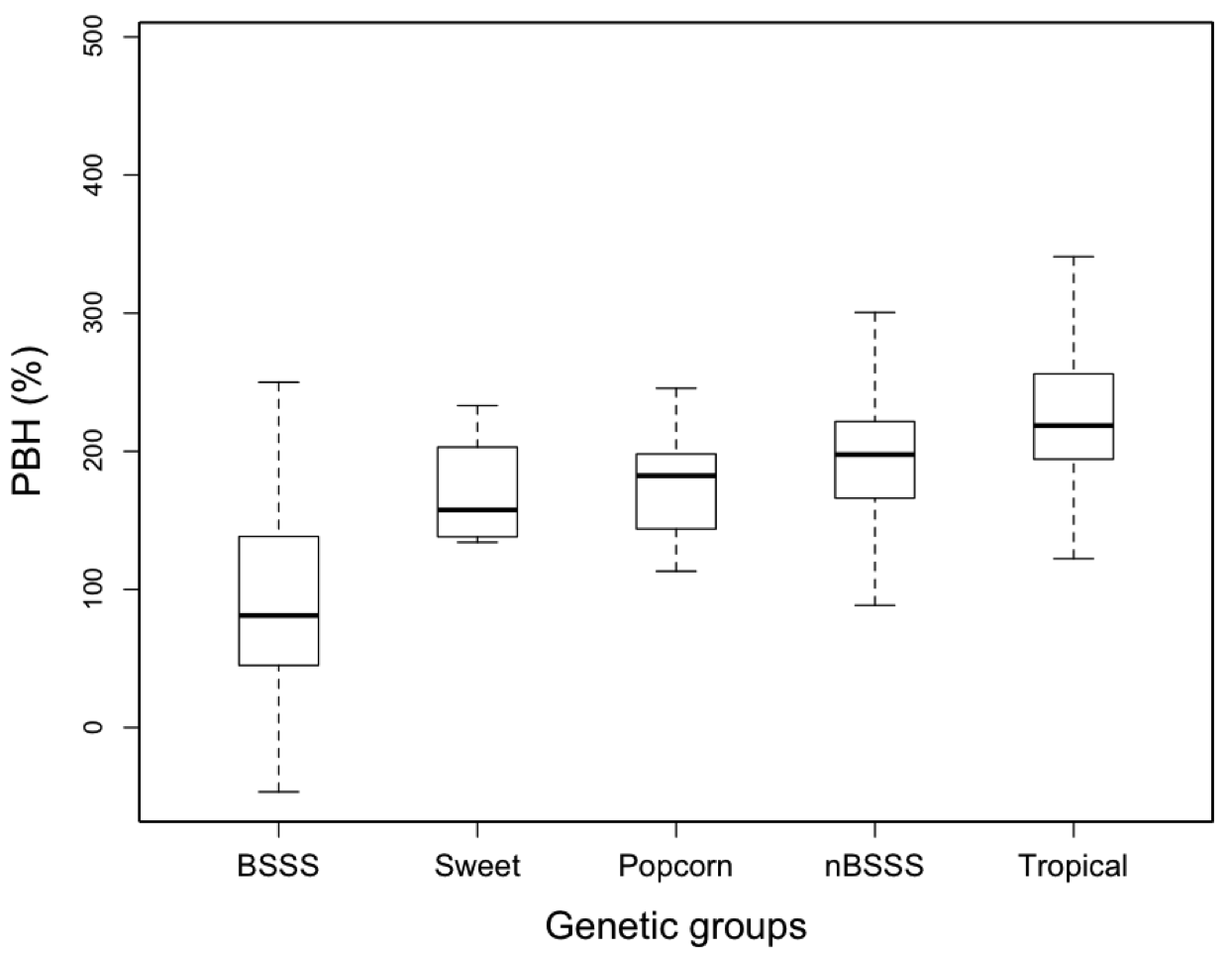}
    \caption{Distribution of best parent heterosis (BPH) for plant yield in population\,A. BSSS and nBSSS indicate the stiff stalk and non-stiff stalk groups.}
   \label{figureS4}
  \end{center}
\end{figure*}

\end{document}